\documentclass[11pt,twoside]{article}
\usepackage{asp2004}
\usepackage{psfig}
\usepackage{epsf}
\usepackage{graphicx}
\usepackage{lscape}

\begin{document}

\title{Imaging the Low Red-shift Cosmic Web}
\author{Robert Braun}
\affil{ASTRON, P.O. Box 2, 7990 AA Dwingeloo, The Netherlands}
\author{David A. Thilker}
\affil{Center for Astrophysical Research, The Johns Hopkins University, \\
3400 North Charles Street, Baltimore, MD 21218 USA}

\begin{abstract}

The first image of a Cosmic Web, Lyman Limit System has just been made in HI
emission within the Local Group filament connecting M31 and M33. The
corresponding HI distribution function is in very good agreement with that of
the QSO absorption lines, confirming the 30-fold increase in surface area
expected between 10$^{19}$ cm$^{-2}$ and 10$^{17}$ cm$^{-2}$. The critical
observational challenge is crossing the ``HI desert'', the range of
log(N$_{HI}$) from about 19.5 down to 18, over which photo-ionization by the
intergalactic radiation field produces an exponential decline in the neutral
fraction from essentially unity down to a few percent. Nature is kinder again
to the HI observer below log(N$_{HI}$)~=~18, where the neutral fraction
decreases only very slowly with log(N$_{HI}$). Average spectra of the M31/M33
filament suggest a kinetic temperature of 2$\times10^5$ K for the parent ion
population, consistent with significant shock-heating. The brightest knot
within the filament, with only a 2$\times10^4$ K apparent temperature, is
suggestive of localized cooling and condensation. We have initiated two
complimentary surveys that should lay the groundwork for a comprehensive study
of the Cosmic Web phenomenon in HI emission. When combined with targeted
optical and UV absorption line observations, the total baryonic masses and
enrichment histories of the Cosmic Web could be determined over the complete
range of environmental over-densities.

\end{abstract}

\section{Introduction}

Extragalactic astronomy has traditionally focused on the regions of extreme
cosmic over-density that we know as galaxies. Only in recent years has the
realization emerged that galaxies do not dominate the universal baryon budget
but are merely the brightest pearls of an underlying Cosmic Web.  Filamentary
components extending between the massive galaxies are a conspicuous prediction
of high resolution numerical models of structure formation
\citep[eg.][]{daveetal99,daveetal01}. Such calculations suggest that in the
current epoch, cosmic baryons are almost equally distributed by mass amongst
three components: (1) galactic concentrations, (2) a warm-hot intergalactic
medium (WHIM) and (3) a diffuse intergalactic medium. These three components
are each coupled to a decreasing range of baryonic over-density:
$log(\rho_{\sc H}/\overline \rho_{\sc H})>3.5$, 1--3.5, and $<$ 1 and are
probed by QSO absorption lines with specific ranges of neutral column density:
$log(N_{HI})~>~18$, 14--18, and $<$ 14. The neutral fraction is thought to
decrease with decreasing column density from about 100\% for 
log(N$_{HI})~\ge~$19.5 to about 1\% at log(N$_{HI}$)~=~17, to less than 0.1\%
at log(N$_{HI}$)~=~13. Although a very wide range of physical conditions can
be found within galaxies, the WHIM is thought to be a condensed shock-heated
phase with temperature in the range 10$^5$--10$^7$~K, while the diffuse
intergalactic medium (IGM) is
predominantly photo-ionized with temperature near 10$^4$~K. A complicating
factor to this simple picture is the growing suspicion that the gas accretion
process, traced by the $log(N_{HI})$~=~14--18 systems, may well occur in two
rather different regimes \citep[eg.][]{binney04,keresetal04}. Low to moderate
mass galaxies (M$_{Vir}~<~10^{12}$M$_\odot$) may experience primarily
``cold-mode'' accretion (T~$\sim$~10$^{4.5}$~K) along filaments, while only
more massive systems may be dominated by the more isotropic ``hot-mode''
accretion (T~$\sim$~10$^{5.5}$~K), which until recently was thought to be
universal \citep[eg.][]{reesostriker77,daveetal01}.

The strongest observational constraints on this picture come from the
statistics of the QSO absorption lines. Enough of such QSO spectra have been
obtained to allow good statistical determinations to be made of the rate of
occurrence of intervening absorbers as function of their column density. By
binning such data in red-shift intervals, it has even been possible to gauge
the cosmic evolution of intervening absorbers
\citep{storrie00}. Inter-galactic space has apparently become
continuously tidier by about an order of magnitude from red-shifts of several
down to zero; with a decreasing cross-section of high column absorbers. At the
current epoch we can now confidently predict that in going down from HI column
densities of 10$^{19}$ cm$^{-2}$ (which define the current ``edges'' of
well-studied nearby galaxies in HI emission) to 10$^{17}$ cm$^{-2}$, the
surface area will increase by a factor of 30. The critical observational
challenge is crossing the ``HI desert'', the range of log(N$_{HI}$) from about
19.5 down to 18, over which photo-ionization by the intergalactic radiation
field produces an exponential decline in the neutral fraction from essentially
unity down to a few percent \citep[eg.][]{doveshull94}. Nature is
kinder again to the HI observer below log(N$_{HI}$)~=~18, where the neutral
fraction decreases only very slowly with log(N$_{HI}$). The baryonic mass
traced by this gas (with a 1\% or less neutral fraction) is expected to be
comparable to that within the galaxies, as noted above.

But how are these low N$_{HI}$ systems distributed and what are their
kinematics?  These are questions which can not be addressed with the QSO
absorption line data. The areal density of suitable background sources is far
too low to allow ``imaging'' of the intervening low column density systems in
absorption. Direct detection of the free-free continuum or recombination line
emission from the ionized gas has also proven well beyond the capabilities of
current X-ray and optical instrumentation \citep[eg.][]{mittazetal04}. For
example, the expected H$\alpha$ emission measure is only about
EM~=~5$\times10^{-4}$ cm$^{-6}$ pc. The very best current H$\alpha$ imaging
results reach down to about EM~=~0.1 cm$^{-6}$ pc, which is still orders of
magnitude removed from what would be needed.

\begin{figure}[!ht]
\resizebox{\hsize}{!}{\includegraphics[width=7.5cm]{braun_cwf1.ps}
  \includegraphics[width=6.5cm]{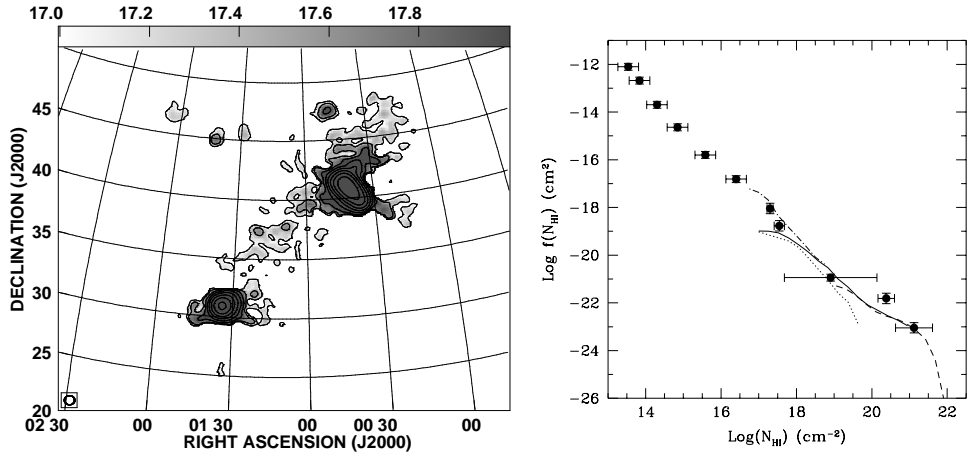}} 
 \caption{{\bf Left:} Integrated HI emission from features which are
 kinematically associated with M31 and M33.  The grey-scale varies between
 log(N$_{HI}$)~=~17~--~18, for N$_{HI}$ in units of cm$^{-2}$. Contours are
 drawn at log(N$_{HI}$)~=~17, 17.5, 18, $\dots$ 20.5. M31 is located at
 (RA,Dec)~=~(00:43,+41$^\circ$) and M33 at (RA,Dec)~=~(01:34,+30$^\circ$), The
 two galaxies are connected by a diffuse filament joining the systemic
 velocities. {Right:} The distribution function of HI column density due to
 M31 and it's environment. The data from three HI surveys of M31 are combined
 in this figure to probe column densities over a total range of some five
 orders of magnitude. The dashed line is from the WSRT mosaic 
 \citep{braunetal04} with 1$^\prime$ resolution over 80$\times$40~kpc, the
 dotted and solid lines from our GBT survey \citep{thilkeretal04}
 with 9$^\prime$ resolution over 95$\times$95~kpc and the dot-dash line from
 the wide-field WSRT survey \citep{braunthilker04} with 49$^\prime$
 resolution out to 150~kpc radius. The filled circles with error-bars are the
 low red-shift QSO absorption line data as tabulated by 
 \citet{corbellibandiera02}. }
\label{fig:m31m33}
\end{figure}

\section{First Imaging Results}

Although conventional imaging in the 21cm emission line of neutral hydrogen
has not typically reached column densities below about 10$^{19}$ cm$^{-2}$,
this is not a fundamental limitation. Long integrations with an
(almost-)filled aperture can achieve the required brightness sensitivity to
permit direct imaging of the small neutral fraction within the Cosmic Web
filaments between galaxies. The first detection of such diffuse filaments in
the extended environment of M31 has just been made by
\citet{braunthilker04}. This was accomplished by utilizing total power
measurements made with the fourteen 25m dishes of the Westerbork Synthesis
Radio Telescope (WSRT). A series of drift-scan observations were used to
obtain Nyquist-sampled HI imaging of a region 60$\times$30 degrees region in
extent, centered approximately on the position of M31.  Although the angular
resolution is low (effective beam of 49~arcmin, corresponding to 11~kpc at the
M31 distance) the column density sensitivity is very high (4$\times10^{16}$
cm$^{-2}$ RMS over 17~km~s$^{-1}$).  A diffuse filament is detected connecting
the systemic velocities of M31 to M33 (at a projected separation of 200 kpc)
and also extending away from M31 in the anti-M33 direction as shown in
Fig.~\ref{fig:m31m33}. This diffuse filament appears to be fueling denser
gaseous streams and filaments in the outskirts of both galaxies.  Peak neutral
column densities within the filament only amount to some 3$\times10^{17}$
cm$^{-2}$. The extremely diffuse nature of the HI has been confirmed by
pointed Green Bank Telescope (GBT) observations of a local peak in the
filament which yields the same low peak column density ($3\times10^{17}$
cm$^{-2}$), despite a telescope beam area that is 25 times smaller.

The interaction zone of the diffuse filament with M31 has been studied in
complimentary surveys: a 6$\times$6 degree field imaged with the GBT
\citep{thilkeretal04} and a 5$\times$2 degree WSRT mosaic of nearly 200
synthesis pointings \citep{braunetal02,braunetal04}. Our three surveys permit
calculation of the HI distribution function from HI emission measurements
(rather than QSO absorption measurements) over an unprecedented range in
log(N$_{HI}$)~=~17.2 to log(N$_{HI}$)~=~21.9 as also shown in
Fig.~\ref{fig:m31m33}. The N$_{HI}$ data for the M31 environment were
normalized to the average space density of galaxies using the HIMF of
\citet{zwaanetal03}.

The HI distribution function of these structures agrees very well with that of
the low red-shift QSO absorption lines which are also plotted in the figure as
filled circles with error bars. The predicted factor of 30 increase in
surface covering factor for low N$_{HI}$ emission has been observationally
verified. In so doing, it has been possible to provide the first 
image of a Lyman Limit absorption system. The morphology and kinematics are
in keeping with the cosmic web hypothesis outlined above, although the region
we have currently probed only extends to about the virial radius of M31. 

\begin{figure}[!ht]
\resizebox{\hsize}{!}{\includegraphics{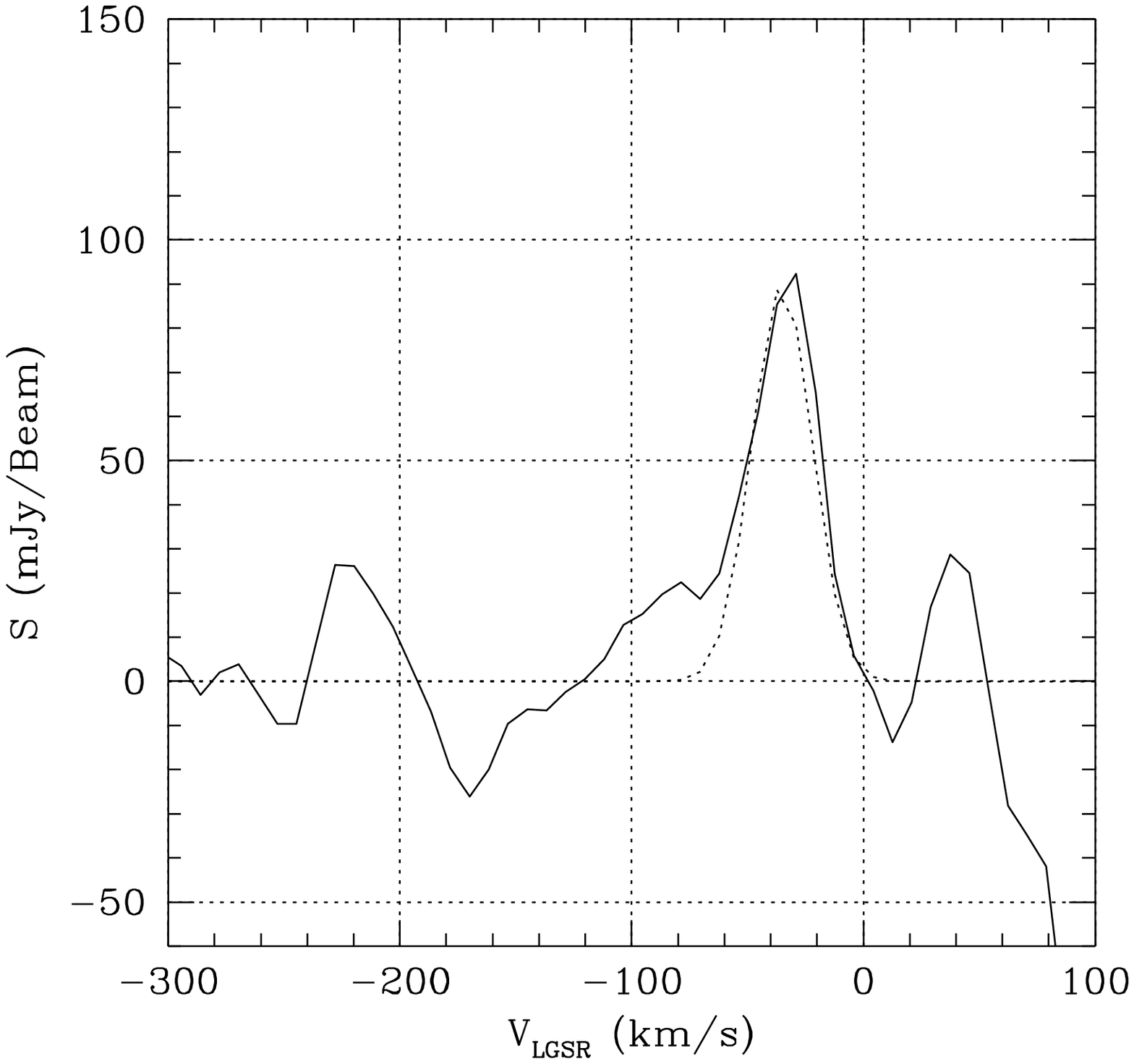} \includegraphics{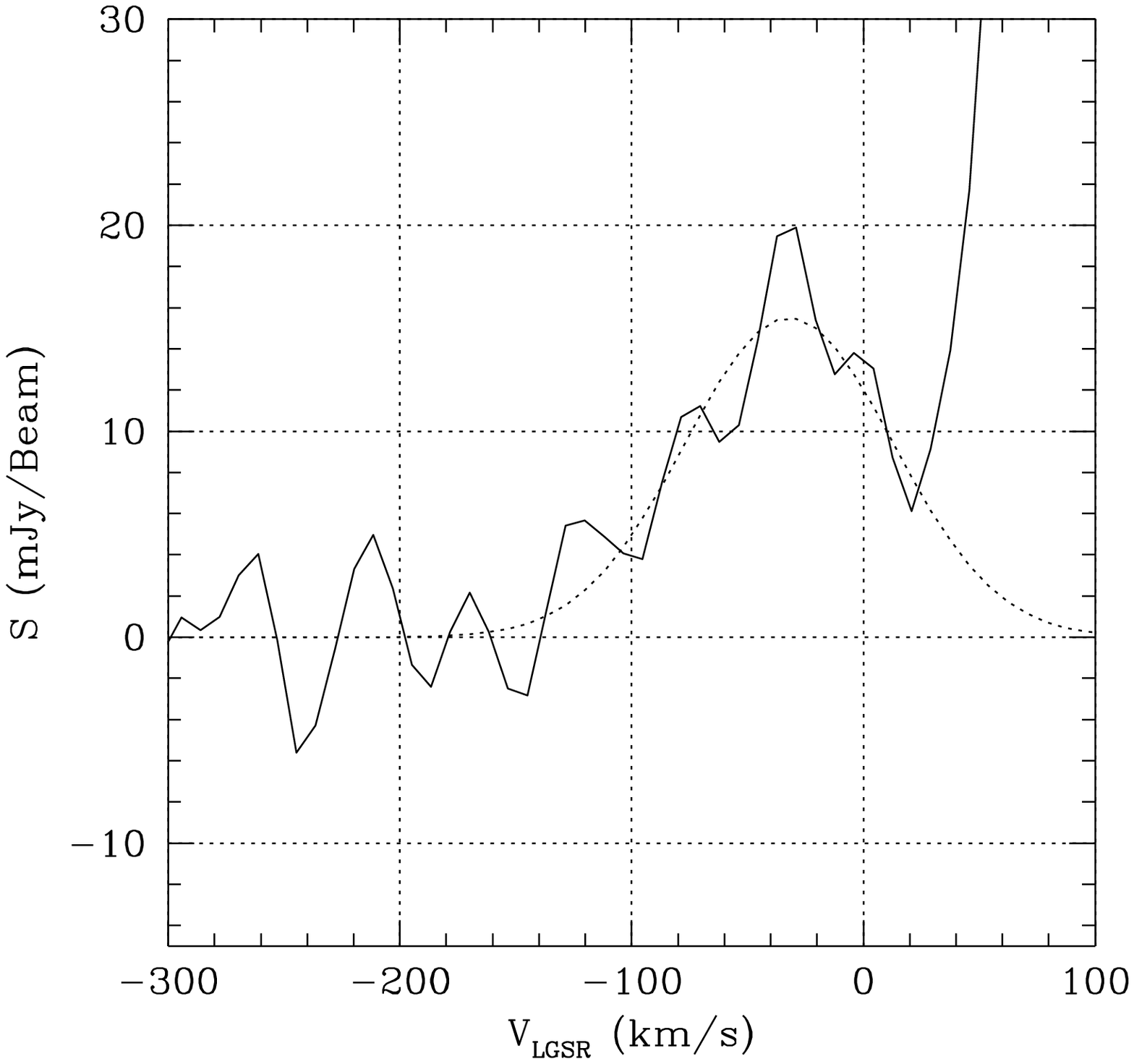}}
 \caption{{\bf Left:} HI spectrum of the brightest clump within the M31/M33
  filament. A Gaussian with 13 km~s$^{-1}$ dispersion, corresponding to a
  thermal linewidth of about 17000~K, is overlaid. {\bf Right:} Average HI
  spectrum of a region of about 4$\times$3 degree within the M31/M33
  filament. A 45 km~s$^{-1}$ dispersion Gaussian, corresponding to about
  2$\times10^5$ K thermal line-width, is overlaid.  }
\label{fig:filspec}
\end{figure}

We have attempted to assess the physical conditions within the M31/M33
filament by extracting and analyzing HI emission spectra of sufficient
signal-to-noise.  The WSRT spectrum toward the brightest knot at
$(\alpha,\delta)_{J2000}~=~(01:20:29.4,+37:22:33)$ is shown in
Fig.~\ref{fig:filspec}. The negative values at positive V$_{LGSR}$ are a
consequence of Galactic foreground emission. A Gaussian with 13 km~s$^{-1}$
dispersion, corresponding to a thermal linewidth of about 17000~K, is
overlaid.  This concentration was also observed with a pointed 30 minute
ON/OFF integration with the GBT. The GBT spectrum confirms a moderately narrow
line core of similar strength, and also suggests an underlying broader
component. In the right hand panel of Fig.~\ref{fig:filspec} we show the
spectrum obtained by averaging over a sigificant fraction of the filament,
corresponding to a box of 4$\times$3 degree extent in (RA,Dec). From the first
moment distribution along the filament \citep{braunthilker04} we can estimate
the RMS contributions of the large-scale velocity gradient and additional
smaller-scale line-of-sight velocity structure to the total linewidth to be
less than about 20~km~s$^{-1}$. This is much smaller than the observed
line-width over the filament, which is well-approximated by a 45~km~s$^{-1}$
dispersion Gaussian, as overlaid in the figure. This line-width, if thermal in
nature, would correspond to a kinetic temperature of 2$\times10^5$ K.

The broad line-width seen in the bulk of the M31/M33 filament is consistent
with the expected kinetic temperature in shock-heated WHIM filaments. While it
may seem counter-intuitive that neutral hydrogen emission could display such a
high kinetic temperature, it must be remembered that the HI may be a trace
constituent of a highly ionized plasma. A small fraction of the individual
hydrogen ions in such a plasma is continuously undergoing recombination. The
emission line-width must then reflect the kinetic temperature of the parent
ion population. Only when the neutral fraction becomes significant, will the
emission line-width reflect the kinetic temperature of the atoms.

\section{New Initiatives}

Two complimentary initiatives have just been begun which should lay the
groundwork for a comprehensive study of the Cosmic Web phenomenon and provide
strong constraints on cosmological models of galaxy evolution.

\begin{figure}[!ht]
\resizebox{\hsize}{!}{\includegraphics{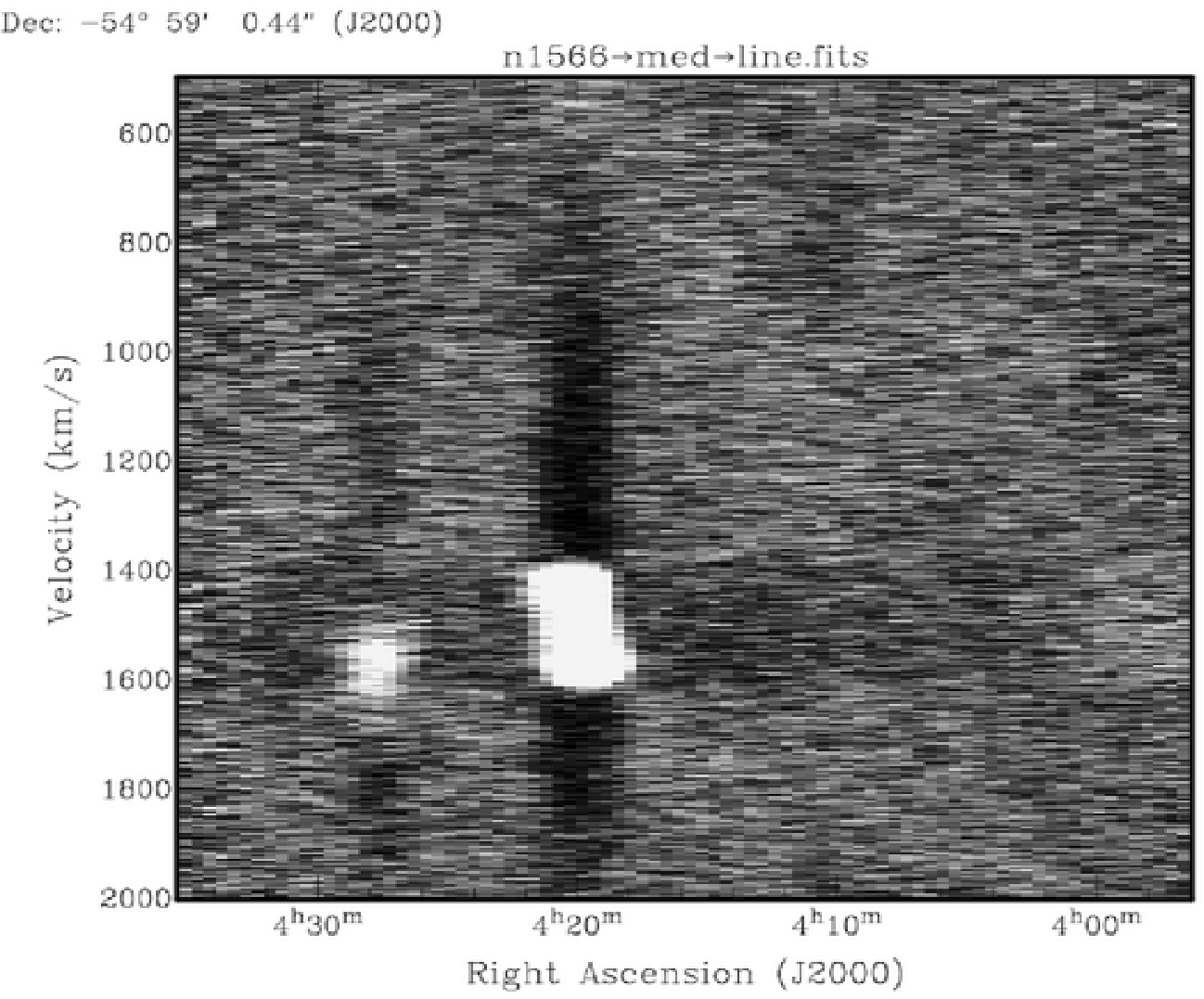} \includegraphics{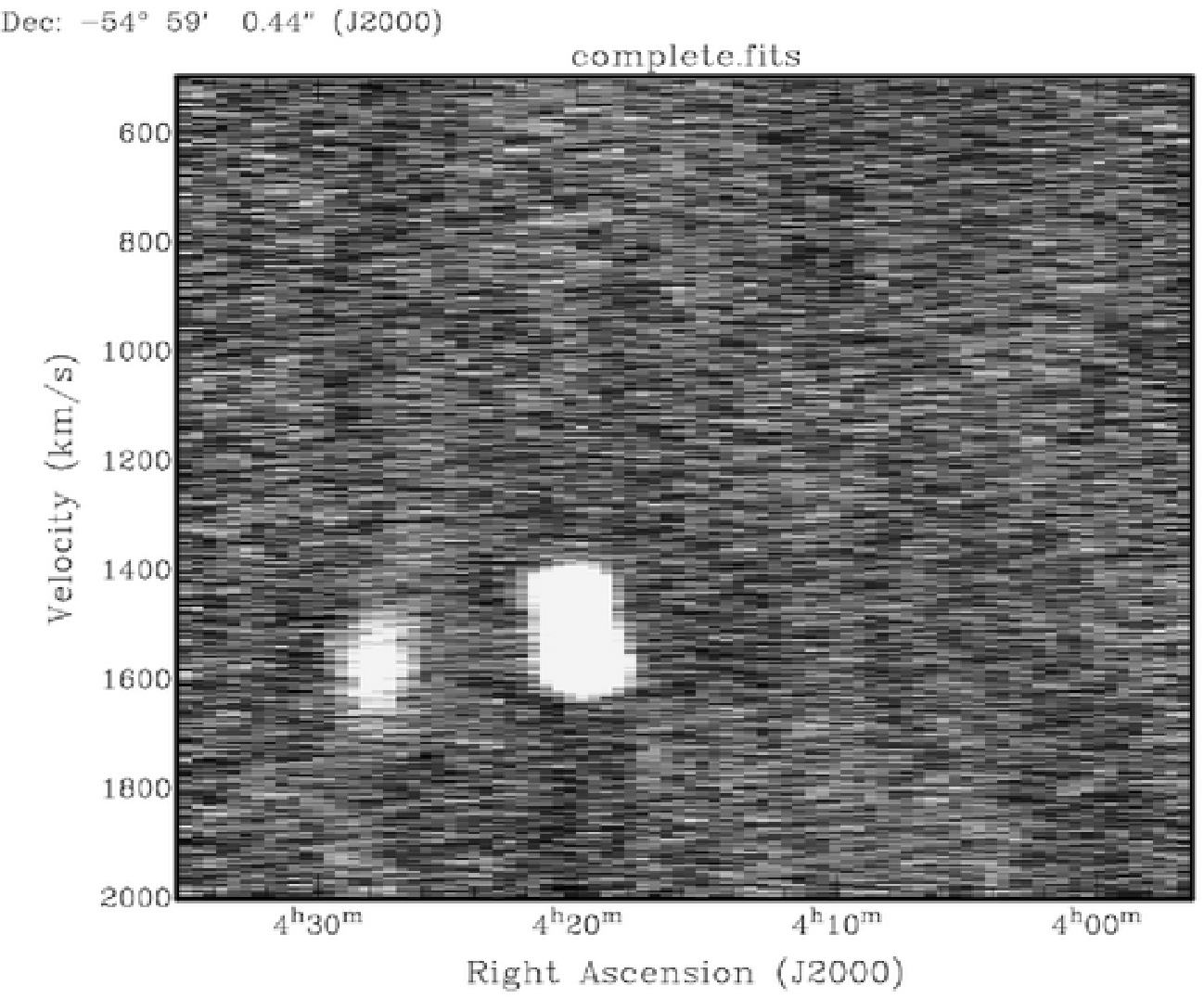}}
 \caption{ Comparison of the current HIPASS data-product (left) with a very
 preliminary re-processed version of the same field (right). Note the severe
 negative artifacts in the current HIPASS products which completely mask any
 faint structures around bright galaxies. With our improved calibration such
 features are beginning to be detected. }
\label{fig:hipass}
\end{figure}

The first initiative involves a complete reprocessing and analysis of the
recently published HI Parkes All-Sky Survey (HIPASS) of the HI sky between
Declinations of $-$90 and $+$20$^\circ$ \citep{barnesetal01}, together with
existing deeper integrations of selected regions (galaxy groups and
clusters). While the column density sensitivity of even the nominal survey is
quite good ($\Delta$N$_{HI}\sim$4$\times10^{17}$ cm$^{-2}$ RMS over
17~km~s$^{-1}$ in a 15~arcmin beam, and better by factors of a few in selected
regions) the adopted method of band-pass calibration has resulted in strong
negative residuals in the vicinity of bright detected sources. Unfortunately,
this type of artifact makes it impossible to detect or even constrain the
anticipated faint HI cosmic web filaments within the current HIPASS products.
This is illustrated in Fig.~\ref{fig:hipass}, where the existing HIPASS
product is contrasted with a very preliminary re-processed version of the same
data. Although not yet optimized, our improved calibration (based on iterative
signal blanking) is already revealing diffuse HI filaments that were not
previously detected.  Further optimization of the calibration algorithm should
enable the theoretical N$_{HI}$ sensitivity to be achieved over the entire
$\sim3\pi$~sr covered by the HIPASS survey to yield a new, RHIPASS, product.

\begin{figure}[!ht]
\resizebox{\hsize}{!}{\includegraphics{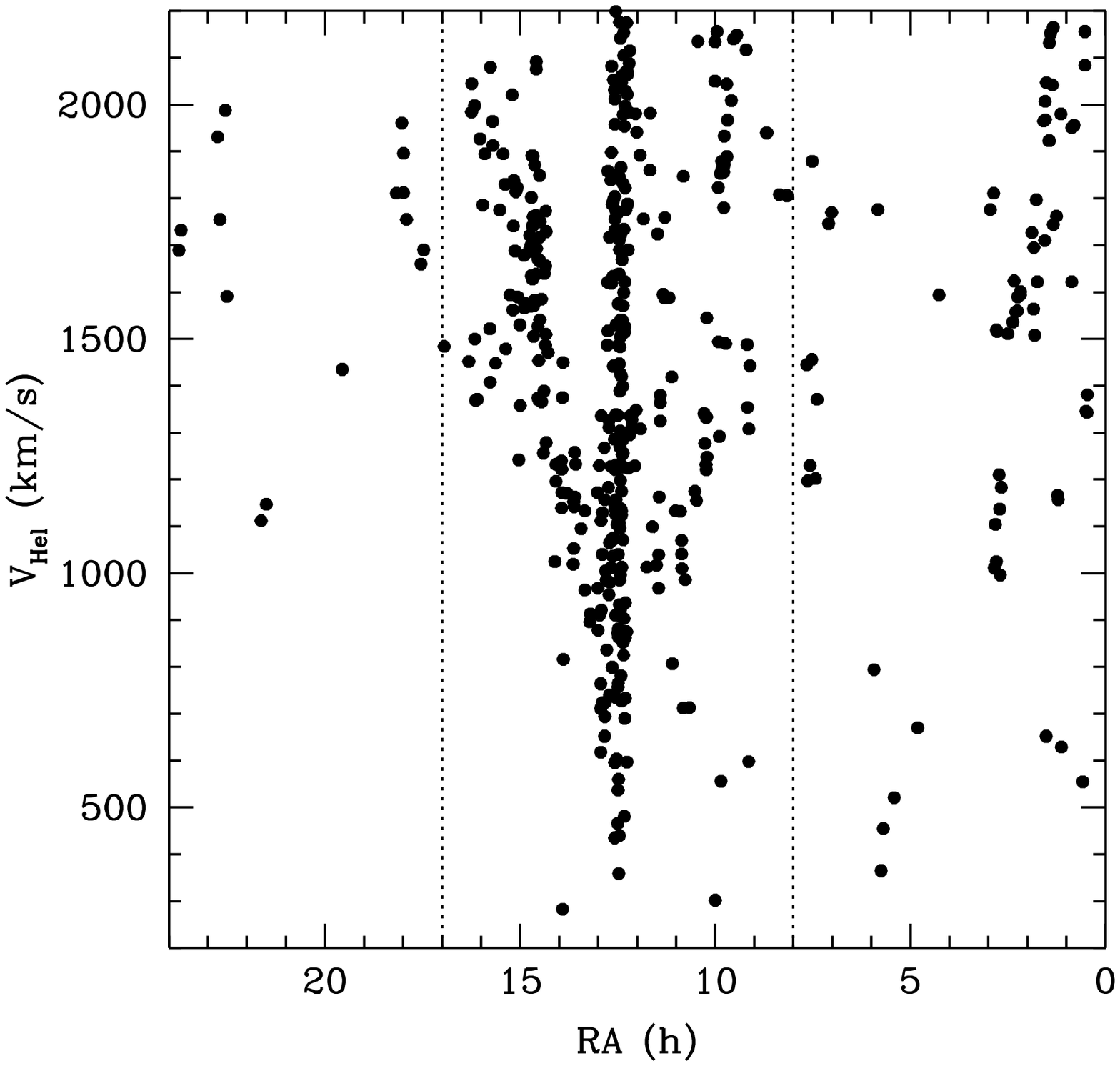} \includegraphics{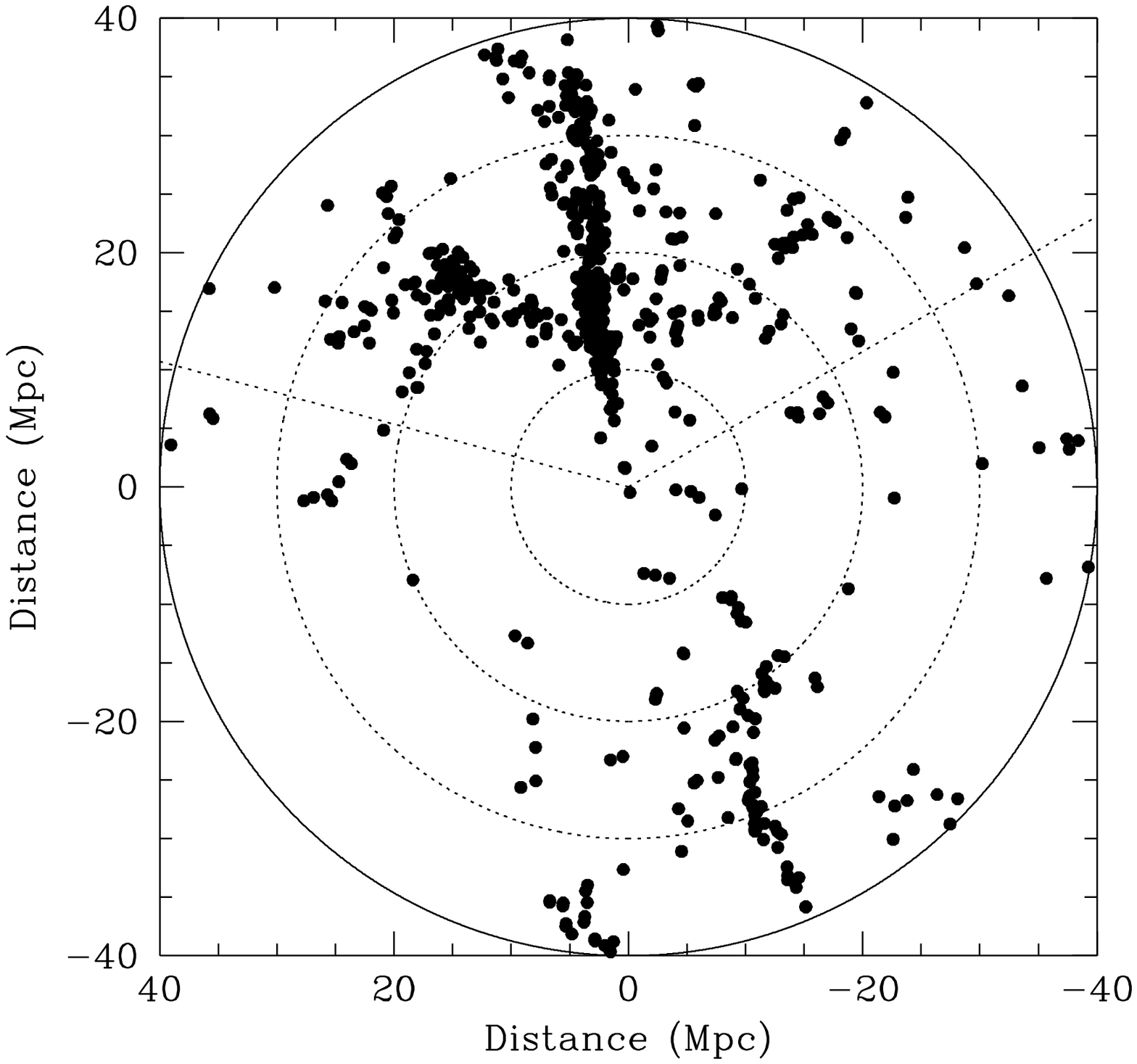}}
 \caption{{\bf Left:} Known galaxies with recession velocity less
 than 2000 km/s in the Declination range $-$1 to +10$^\circ$ are plotted as
 function of RA.  {\bf Right:} Galaxies within 40 Mpc (in the same Dec. range)
 are plotted in a wedge diagram representation based on Virgo-corrected radial
 velocity and H$_0$~=~75~km/s/Mpc. The ongoing WSRT Virgo Filament Survey
 region is delineated with the dotted lines in both panels. }
\label{fig:virfil}
\end{figure}

The second initiative involves a novel and extremely sensitive method for
detection of faint HI emission features which employs the WSRT to simulate a
filled aperture, while retaining the excellent spectral baseline and PSF
properties of an interferometer. The method utilizes the fact that 12 of the
14 WSRT 25~m telescopes can be positioned at regular intervals of 144~m on an
East-West line. When observing low Declination fields at large positive or
negative hour angles, it is possible to form a filled aperture in projection
of 300$\times$25~m extent. Although these geometric conditions are only
satisfied for a few minutes each day (for a given pointing center), even a
single minute of integration already provides a
$\Delta$N$_{HI}\sim$5$\times10^{17}$ cm$^{-2}$ RMS over 20~km~s$^{-1}$ in a
35$\times$3~arcmin beam over a 35$\times$35 arcmin instantaneous
field-of-view. Accumulating integration time over a period of a few weeks
makes it possible to probe the N$_{HI}~\sim~10^{17}$ cm$^{-2}$ regime over
large areas of the sky. Test observations using the WSRT in this observing
mode have already been carried out in March 2004. These tests have verified
both the expected high sensitivity and the excellent spectral baseline and PSF
properties. The specific observing program that will be undertaken is a deep
fully-sampled survey of the galaxy filament joining the Local Group to the
Virgo Cluster, extending from 8 to 17 hours in RA and from $-$1 to
$+$10$^\circ$ in Dec. (as shown in Fig.~\ref{fig:virfil}). Our WSRT Virgo
Filament Survey (WVFS) will probe the extended environment of more than 340
known galaxies covering a wide range of Hubble type and local over-density at
distances between about 2 and 40 Mpc. It is clear from Fig.~\ref{fig:virfil}
that our survey region provides excellent sampling of the Virgo over-density,
the most substantial concentration of matter in the local universe.

The survey has been designed to reach $\Delta$N$_{HI}\sim$2$\times10^{17}$
cm$^{-2}$ over 20~km~s$^{-1}$ in the synthesis data, while the total power
data which is being acquired simultaneously will reach
$\Delta$N$_{HI}\sim$1.3$\times10^{16}$ cm$^{-2}$ for diffuse emission filling
the 35~arcmin primary beam. Survey observations are now scheduled to begin in
December 2004.

We expect many clear detections of diffuse HI filaments to emerge from our
survey. We stress again that the sky area covered at log(N$_{HI}$)~=~17 is
known to be 30 times greater than that currently probed at
log(N$_{HI}$)~=~19. The detected features will be analyzed from the
perspective of their role in on-going galaxy evolution as a function of local
over-density. The observed HI linewidth of the detected features should permit
a straightforward assessment of the relative importance of ``cold-mode''
versus ``hot-mode'' accretion given the very distinctive line-widths
($\sim$30~km~s$^{-1}$ versus $\sim$100~km~s$^{-1}$). Follow-up observations at
higher resolution will be carried out both in HI to study galaxy fueling in
detail as well as optically to look for signs of ongoing (dwarf) galaxy
formation.  Detected filaments will also be correlated with the occurrence of
background AGN, which could serve as probes for absorption measurements in the
optical, UV and X-ray bands. Detection of multiple ionization states for
various species would permit a determination of both the metalicity and
neutral fraction of hydrogen. With even a few of such spot measurements,
estimates of the total mass and current accretion rates could be determined.

\section{Summary}

The first images of Lyman Limit absorption Systems (LLS) have now been
made. These confirm the expected 30-fold increase in sky covering factor
beyond the ``classical'' edges of normal galaxies. Observationally, the LLS
regime is very challenging, since it requires crossing the ``HI desert'', the
range of log(N$_{HI}$) from about 19.5 down to 18, over which photo-ionization
by the intergalactic radiation field produces an exponential decline in the
neutral fraction from essentially unity down to a few percent. However, the
rewards are substantial. A comparable baryonic mass to that currently residing
in galaxies becomes accessible to direct kinematic study. We can now
witness the ongoing formation and evolution of galaxies via accretion along
Cosmic Web filaments. The first indications from the M31/M33 filament are that
much of the detected HI may be associated with a highly ionized plasma of
kinetic temperature of about 2$\times10^5$ K, implying significant shock
heating. Only the brightest knot in the filament appears to be about
2$\times10^4$ K, perhaps indicating ongoing cooling and condensation within
the filament.

Achieving the necessary log(N$_{HI}$)~=~17 sensitivity in HI emission for LLS
imaging studies requires observations with an (almost-)filled aperture. We are
currently pursuing two complimentary initiatives to lay the groundwork for a
comprehensive study of the Cosmic Web phenomenon. In the first we are planning
a complete reprocessing of the HIPASS total power database, optimized for
detection of diffuse filamentary structures near bright galaxies. In the
second we are beginning a wide-field synthesis survey of the galaxy filament
joining the Local Group with the Virgo cluster. In the latter case, we are
utilizing the regular spacing of the WSRT array telescopes to form a filled
aperture in projection of 300$\times$25~m extent, while retaining the
excellent spectral baseline and PSF properties of an interferometer. In this
way we hope to provide the strongest possible constraints on cosmological
models of galaxy evolution, using existing instrumentation. Much deeper
HI imaging with even higher angular resolution will be possible with the
planned SKA \citep{braun04} which should enable resolved Cosmic Web studies
throughout the local universe.

\end{document}